\documentclass[twocolumn,english,preprintnumbers,amsmath,amssymb]{revtex4}
\usepackage[T1]{fontenc}
\usepackage[latin9]{inputenc}
\usepackage{color}
\usepackage{amstext}
\usepackage{graphicx}
\usepackage{setspace}
\usepackage{amssymb}
\usepackage{esint}

\makeatletter
\@ifundefined{textcolor}{}
{%
 \definecolor{BLACK}{gray}{0}
 \definecolor{WHITE}{gray}{1}
 \definecolor{RED}{rgb}{1,0,0}
 \definecolor{GREEN}{rgb}{0,1,0}
 \definecolor{BLUE}{rgb}{0,0,1}
 \definecolor{CYAN}{cmyk}{1,0,0,0}
 \definecolor{MAGENTA}{cmyk}{0,1,0,0}
 \definecolor{YELLOW}{cmyk}{0,0,1,0}
 }

\usepackage{dcolumn}
\usepackage{bm}

\makeatother

\usepackage{babel}

\begin{document}

\title{Statistical properties of random matrix product states}

\author{Silvano Garnerone}

\email{garneron@usc.edu}

\affiliation{Department of Physics and Astronomy and Center for Quantum Information
Science \& Technology, University of Southern California, Los Angeles,
CA 90089 }

\author{Thiago R. de Oliveira}

\affiliation{Department of Physics and Astronomy and Center for Quantum Information
Science \& Technology, University of Southern California, Los Angeles,
CA 90089 }

\author{Stephan Haas}

\affiliation{Department of Physics and Astronomy and Center for Quantum Information
Science \& Technology, University of Southern California, Los Angeles,
CA 90089 }

\author{Paolo Zanardi}

\altaffiliation[Also at ]{Institute for Scientific Interchange, Viale Settimio Severo 65, I-10133 Torino, Italy}

\affiliation{Department of Physics and Astronomy and Center for Quantum Information
Science \& Technology, University of Southern California, Los Angeles,
CA 90089 }

\date{\today}
\begin{abstract}
We study the set of random matrix product states (RMPS) introduced
in PRA \textbf{81}, 032336 as a tool to explore foundational aspects of quantum
statistical mechanics. In the present work, we provide an accurate
numerical and analytical investigation of the properties of RMPS.
We calculate the average state of the ensemble in the non-homogeneous
case, and numerically check the validity of this result. We also suggest
using RMPS as a tool to approximate properties of general quantum
random states. The numerical simulations presented here support the
accuracy and efficiency of this approximation. These results suggest
that any generalized canonical state can be approximated with high
probability by the reduced density matrix of a random MPS, if the
average MPS coincide with the associated microcanonical ensemble.
\end{abstract}

\pacs{Valid PACS appear here}

\maketitle

\section{Introduction}

The principle of equal a priori probabilities, which is at the foundation
of statistical mechanics, postulates subjective ignorance at the core
of our understanding of macroscopic systems. It assumes that an isolated
system is described by the microcanonical ensemble: an equal mixture
over all possible microscopic configurations. On the other hand, quantum
mechanics provides us with a powerful theory for understanding microscopic
systems. It states that an isolated system is described by a pure
state, which leaves no room for subjective ignorance. In the attempt
to provide a consistent foundation of quantum statistical mechanics
a purely quantum mechanical explanation of the effectiveness of statistical
ensembles has been debated for quite a long time.

Recently an attempt to provide an alternative foundation to statistical
mechanics was suggested, in which entanglement is viewed at the origin
of generalized canonical ensembles. In this approach, there is no
need to assume the principle of equal a priori probability, and the
isolated system can be in a pure state, consistent with quantum mechanics
\cite{PoShWi,GoLeTu}. The mathematical justification for the effectiveness
of this approach is based on the concentration-of-measure phenomenon,
which has also appeared in the literature under the name of typicality
\cite{Leb1}. Simplifying the results of \cite{PoShWi,GoLeTu}, one
can prove that the vast majority of pure states picked uniformly at
random in a sufficiently large Hilbert space will be almost indistinguishable
at the level of sufficiently small subsystems, stemming from the appropriate
statistical mechanical ensemble. Therefore it is not necessary to
assume subjective ignorance over the state of the system, describing
it with a mixed state, since any pure state will typically give very
similar results.

Despite the elegance and simplicity of this approach there are aspects
which need to be addressed in order to understand its concrete effectiveness.
For example, it has already being pointed out in other contexts \cite{EmWeSa,Low}
that it is extremely inefficient to reach or create a random pure
state in a large system. This is due to the fact that the uniformly
distributed random unitary generating the state requires an exponential
number of parameters to be specified, and any dynamical process corresponding
to it will take an exponentially long time. This is in contrast with
the relatively fast process of equilibration, which naturally leads
the subsystem to its generalized canonical ensemble. Therefore, if
typicality has to provide an explanation for the statistical mechanical
ensembles corresponding to equilibrium states, then it should also
be able to account for fast equilibration processes. One way of solving
this problem is to restrict the class of allowed states in the isolated
system. This kind of approach has also been considered in Ref.\cite{BrMoWi,GrFlEi,Low}.

In previous work, we reported first results on the occurrence of typicality
in random matrix product states \cite{GadeOZa}. Constraining the
states of the system to be MPS-like is one way to avoid the efficiency
problem, mentioned above, in the realization of the concentration-of-measure
phenomenon. MPSs are an example of physically meaningful states. They
can be ground states of gapped Hamiltonians \cite{Has,VeCi}, and
they are at the core of some computational techniques, based on quantum
information theory, which have been recently introduced \cite{Hal}.
The principal reason for the effectiveness of MPS is that the ground
states of many relevant systems can be well approximated by them (see
\cite{VeCiMu} for a recent review and original references). Here
we present more details about typicality in MPSs giving further evidence
of its relevance. 

In Sec. II we provide some background on the literature concerning
typicality in random quantum states. In Sec. III we define the particular
set of random MPS we are interested in and review previous results
on the concentration-of-measure phenomenon in this context \cite{GadeOZa}.
In Sec. IV we provide a detailed numerical analysis of the initial
results discussed in Ref. \cite{GadeOZa} and present an analytical
derivation of the average non-homogeneous RMPS. We also study specific
statistical properties of RMPS and show how they can be used to efficiently
approximate some characteristics of general random pure states. Sec.
V is devoted to conclusions.

\section{Typicality Background}

Arguments related to typicality have been presented throughout the
quantum and statistical mechanics literature since the very beginning
of this field (e.g. \cite{Sch} and chapter 6 in \cite{GeMiMa}).
More recently, Lebowitz has emphasized the importance of typicality
in classical statistical mechanics in the context of the second law
of thermodynamics \cite{Leb2}. In Refs. \cite{PoShWi,GoLeTu} the
importance of typicality was recognized also in the context of quantum
statistical mechanics. A different and more heuristic approach was
used in \cite{Rei1}. A nice review with original contributions can
be found in \cite{GeMiMa}. In the following, we provide a brief overview
of some of these works.

The work in \cite{PoShWi} makes direct use of the concentration-of-measure
phenomenon and focuses on the properties of the subsystem's states.
The concentration-of-measure phenomenon is a well known topic in the
mathematical literature and appears in many different contexts \cite{MiSc,Tal,Led}.
Roughly speaking, suppose we have a function $f:V^{n}\rightarrow\mathbb{C}$,
whose domain is an n-dimensional vector space equipped with a probability
measure. If $f$ does not oscillate too much then, for $n$ sufficiently
large, $f$ is almost constant, with very high probability. More formally:
\begin{equation}
\text{Prob}\left\{ |f-\bar{f}|\geq\epsilon\right\} \leq k_{1}exp(-k_{2}\epsilon^{2}n/\eta^{2}),\end{equation}
where $\bar{f}$ is the average value of $f$, $k_{1,2}$ are some
universal positive constants, and $\eta$ is the Lipschitz constant
of $f$. In this sense $f$ is concentrated around its average value.
The Lipschitz constant can be defined as the supremum of the gradient.
Within this formalism one can prove, for example, that the area of
a hypersphere concentrates around the equator \cite{MiSc}. When the
domain of $f$ is a hypersphere the concentration-of-measure phenomenon
is also referred to as Levi's lemma \cite{MiSc}.

\subsection{Local typicality}

\textcolor{black}{Suppose we have a system $R$, whose Hilbert space
$\mathcal{H}_{R}$ belongs to }$\mathcal{H}_{S}\otimes\mathcal{H}_{B}$,
the tensor product of a subsystem $S$ and a bath $B$. We denote
the dimensions of the corresponding Hilbert spaces with $d_{R},\, d_{S}$
and $d_{B}$ respectively. The states in the isolated system $R$
may satisfy some constraints or restrictions, for example belonging
to a small energy-shell with respect to a Hamiltonian. We now consider
a uniformly random state $|\psi\rangle\in\mathcal{H}_{R}$. This can
be achieved by generating $|\psi\rangle$ with a random unitary matrix,
distributed according to the Haar measure, acting on some reference
state in $\mathcal{H}_{R}$. Consider now the state of the subsystem
$\rho_{S}=\text{Tr}_{B}|\psi\rangle\langle\psi|$. The results in
\cite{PoShWi} show that, for almost all $|\psi\rangle$, $\rho_{S}$
will be very close in trace norm to the generalized canonical distribution.
This is proved by defining the function $D_{1}\equiv||\rho_{s}-\overline{\rho_{s}}||_{1}$,
for which the Lipschitz constant is 2, and using Levi's lemma to obtain
a concentration-of-measure result on the function $D_{1}$. Then one
has to show that also the average value $\overline{D_{1}}$ is small:
most states are almost at the same distance from the average state,
and this distance is small. This last result comes from the following
inequalities \cite{PoShWi} \begin{equation}
\overline{D}_{1}\leq\sqrt{d_{S}\text{Tr}[\overline{\rho}_{B}^{2}]}\leq\sqrt{d_{S}^{2}/d_{R}}.\end{equation}

From this, using Chebyshev's inequality and the fact that $D_{1}\geq0$,
one obtains a first concentration result, i.e. a bound on the probability
of fluctuations. However, Levi's lemma gives a much stronger bound
\cite{PoShWi}

\begin{equation}
\text{Prob}\left[D_{1}\geq\epsilon+\sqrt{d_{S}\text{Tr}[\overline{\rho}_{B}^{2}]}\right]\leq2e^{-Cd_{R}\epsilon^{2}},\end{equation}
with $(C=18\pi^{3})^{-1}$. Note that the probability is exponentially
suppressed in the dimension of the Hilbert space and not just in the
size of $R$, which would be the result obtained using Chebyshev's
inequality.

From the bound on the state distance one can easily derive a similar
result for any observable, in this way establishing typicality for
the subsystem. It is worth emphasizing that this derivation of a generalized
canonical ensemble is fully consistent with the quantum formalism
and makes no use of the principle of equal a priori probability. On
the other hand, the entanglement between the subsystem and the bath
is seen as the main mechanism through which incomplete knowledge over
the state of the subsystem emerges.

\subsection{Global typicality}

\textcolor{black}{An alternative approach to typicality \cite{Rei1}
aims to establish it at the level of the isolated global system, and
to generalize it further to probability distributions other than uniform.
Let us consider a basis decomposition of a random state\begin{equation}
|\psi\rangle=\sum_{n}c_{n}|n\rangle,\end{equation}
where the coefficients $c_{n}$ are random. We make the following
two assumptions: $c_{n}$ are independent (but do not need to be identically
distributed) random variables, and they are phase independent, i.e.
}\begin{equation}
\mathcal{P}(\psi)\to\mathcal{P}(c_{1},c_{2},...)=\prod_{n}\mathcal{P}(|c_{n}|).\end{equation}
This kind of probability distribution gives an average state which
is diagonal and coincides with the dephased state \cite{Rei1}. Defining
$\rho\equiv|\psi\rangle\langle\psi|$, and assuming that the purity
of the average state is low ($\text{Tr}\left[\bar{\rho}^{2}\right]\ll1$),
one can show that the variance in the expectation value of $A=\langle\psi|A|\psi\rangle$
is small \begin{equation}
\sigma_{A}^{2}\equiv\overline{(A-\bar{A})^{2}}\leq||A||_{\infty}^{2}(\max_{n}q_{n})\text{Tr}[\bar{\rho}^{2}]\end{equation}
with $q_{n}$ the normalized variance of the random variables $|c_{n}|^{2}$.
Note that the operator norm $||A||_{\infty}^{2}$ quantifies the range
of possible values for the observable $A$. From the above inequality
it follows that a random pure state is likely to yield expectation
values for $A$ which are very close to the ensemble average $\text{Tr}\left[\bar{\rho}A\right]$.
This behavior occurs for probability distributions such that $\text{Tr}\left[\bar{\rho}^{2}\right]\ll1$,
and for observables, where $||A||_{\infty}^{2}$ does not increase
linearly with the dimension of the Hilbert space (which can be argued
to be the physically most interesting, see also the comments
in \cite{Bro}). A trivial way to guarantee that $||A||_{\infty}^{2}$
does not increase is to restrict to local observables, supported on
finitely many subsystems. From the bound on the variance one can use
Chebyshev's inequality to bound the probability that $A$ will be
far way from its average value $\overline{A}$.

\subsection{Remarks}

\textcolor{black}{The use of non-uniform probability distributions
for the random state of the isolated system could be considered unphysical,
since they are not invariant under a unitary transformation and consequently
they depend on the chosen basis. But, in Ref. \cite{Rei1} it was
argued that also assuming that all eigenstates are populated with
the same probability is rather unphysical. In this sense, these results
are interesting since they show that typicality does not depend on
the details of the probability distribution. The canonical ensemble,
for example, can be recovered under much more general and realistic
conditions than the ones usually considered, such as the principle
of equal a priori distribution. Another advantage of this approach
is that it establishes typicality at a global level. From this it
follows that typicality is a concept independent of entanglement.
Typicality at the subsystem level follows naturally from typicality
at the global level. However, since there are some observables for
which typicality does not hold at the global level, it would be interesting
to understand their physical relevance.}

\textcolor{black}{It is important to underline that the above discussion
does not show thermalization emerging from a dynamical process. Nevertheless,
they show that most of the states of the Hilbert space are thermalized.
Therefore one expects that most of the dynamics will give thermalization,
the exception being some particular dynamics that keep the states
within an exponentially small atypical subspace.} In fact, in Refs.
\cite{Rei2,Linden} it has been shown in great generality that most
of the time the state is close to some fixed state, the time averaged
one, and in this sense there is equilibration of the system. An interest\textcolor{black}{ing
area of research in this respect is the unitary dynamics of isolated
systems after a sudden quench \cite{RiDuOl,CrDaEi}.}

The typicality arguments reviewed above provide a consistent foundation
for quantum statistical mechanics, although some questions can still
be raised. In particular, one can argue about the exponential amount
of resources necessary to generate uniformly distributed random states
and the fact that apparently \textcolor{black}{no} dynamical process
can generate such a state efficiently. In this sense, one might also
wonder whether most of the states in the Hilbert space are physically
realizable. \textcolor{black}{These arguments are also relevant in
condensed matter, where one is not able to use all possible states
because of the exponential number of parameters that would be needed.
Due to the locality of the interactions only a subset of states will
be effectively useful in the description of the system. The problem
then is to find an efficient representation for such physical states.}

\textcolor{black}{Therefore it is important to study if typicality
is still valid for a smaller set of more relevant states. Hence, we
decide to focus on MPSs, relying on their importance for Hamiltonians
with local interactions.}

\section{Random matrix product states}

A matrix product state is a pure quantum state whose coefficients
are specified by a product of matrices. For the case of periodic boundary
conditions (PBC) a MPS can be written as \begin{equation}
\sum_{i_{1},...,i_{N}}{\rm Tr}\left(A^{i_{1}}[1]\cdots A^{i_{N}}[N]\right)|i_{1}\cdots i_{N}\rangle,\end{equation}
whereas with open boundary conditions (OBC) one has \begin{equation}
\sum_{i_{1},...,i_{N}}\langle\phi_{I}|A^{i_{1}}[1]\cdots A^{i_{N}}[N]|\phi_{F}\rangle|i_{1}\cdots i_{N}\rangle,\end{equation}
with $|\phi_{\{I,F\}}\rangle$ specifying the states at the boundaries,
and $|i_{k}\rangle$ is a local basis at site $k$. The matrices $\{A^{1}[k],A^{2}[k],\dots,A^{D}[k]\}$,
with $k\in\{1,\dots,N\}$, are $\chi$-dimensional complex matrices,
where $D$ is the local Hilbert space dimension. We denote as homogeneous
MPSs the states for which the set $\{A^{1}[k],A^{2}[k],\dots,A^{D}[k]\}$
is the same for all sites $k$. By definition, a MPS is specified
by the set $\{A^{1}[k],A^{2}[k],\dots,A^{D}[k]\}$. However there
may be different sets of matrices that originate the same MPS. In
\cite{PeVeWo} it was shown that this gauge degree of freedom can
be fixed using a canonical form. \textcolor{black}{The fundamental
parameter characterizing the properties of MPS states is $\chi$,
the size of the $A$-matrices. Note that any non-homogeneous MPS is
parametrized by $ND\chi^{2}$ numbers, which can be much less than
the $D^{N}$ values needed for a general state. Furthermore one can
prove that the maximum entanglement a subsystem can have with its
environment depends on $\chi$. It can also be shown that any state
can be described as a MPS for sufficiently large $\chi\propto D^{N}$,
but there is no advantage in such a representation. In \cite{VeCi}
it was proved that the ground state of any one-dimensional local Hamiltonian
is well approximated by a MPS. At the computational level, MPSs are
also very useful in algorithms based on the density matrix renormalization
group \cite{VeCiMu}. For all these reasons MPSs can be considered
to be good representations of physical states of one-dimensional systems,
and it is also possible to generalize this formalism to higher-dimensional
system \cite{VeCiMu}.}

\begin{figure}
\includegraphics[scale=0.4]{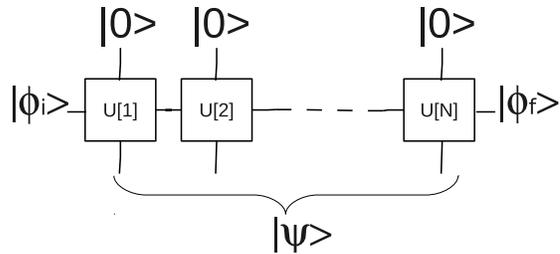}

\caption{Sequential generation of a MPS $|\psi\rangle.$}

\label{fig:fig1}
\end{figure}

In the present work, using the sequential generation of MPSs introduced
in \cite{ScHaWo,PeVeWo}, we consider random MPSs with a clear operational
definition. Consider a quantum spin chain initially in a product state
$|0\rangle^{\otimes N}\in\mathcal{H}_{B}^{\otimes N}$ (with $\mathcal{H}_{B}\simeq\mathbb{C}^{D}$)
and an ancillary system in the state $|\phi_{I}\rangle\in\mathcal{H}_{A}\simeq\mathbb{C}^{\chi}$.
Let $U[k]$ be a unitary operation on $\mathcal{H}_{A}\otimes\mathcal{H}_{B}$,
acting on the ancillary system and the $k$'th site of the chain (see
Fig. \ref{fig:fig1}). The $A[k]$ matrices are then defined by \begin{equation}
A_{\alpha,\beta}^{i}[k]\equiv\langle i,\alpha|U[k]|0,\beta\rangle,\label{eq:Amat}\end{equation}
 where the Greek indices refer to the ancilla space and the Latin
indices to the physical space (see \cite{ZySo,BrCaSo} for related
works on truncated random unitaries). For homogeneous MPSs the index
$k$ is removed, implying that the unitary interaction is the same
for all sites in the spin chain. Due to unitarity we have $\sum_{i}A^{i}[k]^{\dagger}A^{i}[k]=\mathbb{I}_{\chi}$
for all $k$ in the bulk (see \cite{GadeOZa} for more details). This
property, together with a proper normalization of the boundaries,
corresponds to a MPS of unit-norm. Letting the ancilla interact sequentially
with the $N$ sites of the chain and assuming that the ancilla decouples
in the last step (this can be done without loss of generality, as
shown in \cite{ScHaWo}), the state on $\mathcal{H}_{B}^{\otimes N}$
is described by \begin{equation}
|\psi\rangle=\sum_{i_{1},\dots,i_{N}}\langle\phi_{F}|A^{i_{N}}\cdots A^{i_{1}}|\phi_{I}\rangle|i_{N}\cdots i_{1}\rangle,\end{equation}
which is a homogeneous MPS with open boundary conditions. It can be
proved \cite{ScHaWo,PeVeWo} that the set of states generated in this
way is equal to the set of OBC-MPSs. We choose the interaction characterizing
the RMPS ensemble to be represented by random unitary matrices $U[k]$
distributed according to the Haar measure.

Since any state can be described by a MPS when $\chi\propto D^{N}$,
typicality naturally holds true for MPSs with this scaling of $\chi$.
What is not obvious is whether one can have typicality in the set
of RMPSs where $\chi$ increases at most polynomially with the number
of particles $\chi\propto N^{a}$, with some constant $a>0$. In \cite{GadeOZa},
using a concentration-of-measure result for the group of unitary matrices
\cite{MiSc}, we were able to prove typicality for the expectation
values of local observables over $L$ sites \begin{equation}
O\equiv\left(\bigotimes_{k=1}^{L}O[k]\right)\left(\bigotimes_{k=L+1}^{N}\mathbb{I}[k]\right),\end{equation}
 with respect to normalized RMPS $|\psi\rangle$ generated according
to the sequential construction. Let us define \begin{equation}
f\equiv\langle\psi|O|\psi\rangle=Tr\left(\prod_{k=1}^{L}E_{O[k]}\prod_{k=L+1}^{N}E_{\mathbb{I}[k]}\right),\label{eq:f_def}\end{equation}
where the transfer operator $E$ is defined as \begin{equation}
E_{O[k]}\equiv\sum_{i_{k},j_{k}=1}^{D}\langle i_{k}|O[k]|j_{k}\rangle A^{i_{k}}[k]\otimes A^{j_{k}}[k]^{*},\end{equation}
\begin{equation}
E_{\mathbb{I}[k]}\equiv\sum_{i_{k}=1}^{D}A^{i_{k}}[k]\otimes A^{i_{k}}[k]^{*}.\end{equation}
 Since the $A-$matrices are sub-blocks or random unitaries $U$,
they define a random variable $f:U(\chi D)\rightarrow\mathbb{R}$
satisfying \cite{GadeOZa} ($c_{1,2}$ are positive constants) \begin{equation}
\textrm{Pr}\left[\left|f-\overline{f}\right|\geq\epsilon\right]\leq c_{1}\exp\left(-c_{2}\epsilon^{2}D\frac{\chi}{N^{2}}\right).\label{eq:com_f}\end{equation}
 This means that if $\chi\propto N^{a},$ with $a>2,$ increasing
the size of the system renders the expectation values more concentrated
around their averages, i.e. statistical fluctuations will be suppressed
\footnote{Note that in the case of non-homogeneous RMPS the domain of the function
$f$ is $N\chi D-$dimensional, implying a concentration result for
$\chi\propto N^{a}$and $a>1.$ %
}. From this result one can also derive a weaker concentration result
for the probability of fluctuation of the trace distance at the sub-system
level. Defining $\rho_{s}\equiv Tr_{Env}|\psi\rangle\langle\psi|$,
and analogously for the average state, we have (see Appendix C in
\cite{GadeOZa} for a derivation) \begin{equation}
\textrm{Pr}\left[\Vert\rho_{s}-\overline{\rho_{s}}\Vert_{1}\geq4^{3L/2}\epsilon\right]\leq4^{L}c_{1}exp(-c_{2}\epsilon^{2}D\frac{\chi}{N^{2}}).\label{eq:com_state}\end{equation}

This shows that although the number of MPS is much smaller than the
total number of states, picking random matrix product states $|\psi\rangle$
according to the above construction still provides a concentration-of-measure
result for suitable random variables defined through $|\psi\rangle.$
This fact could be exploited numerically for the efficient simulation
of quantum systems. 
Ref. \cite{Whi} provides a possible indication of the role 
that the concentration of measure
may  play in numerical simulations. In this work a different class of
random MPS is constructed and used for the efficient sampling
of local observables. We suspect that the efficiency of the sampling
is due to a concentration of measure phenomenon.

We point out that the above result
concerning the subsystem state is weaker than the one obtained in
\cite{PoShWi}, where the probability of fluctuations is exponentially
suppressed in the dimension of the total Hilbert space, i.e. doubly-exponentially
in $N$. Similar weaker bounds have been obtained in the context of
k-designs \cite{Low}.

\section{Statistical properties}

\subsection{Haar distributed random states and RMPS}

In this section, we discuss numerical signatures of typicality for
the case of random pure states distributed according to the Haar measure.
These simulations are useful in illustrating typicality and the tightness
of the theoretical bounds, and they also serve as a reference point
for understanding the properties of RMPS.

\begin{figure}[htp]
\centering{}\includegraphics[clip,scale=0.4]{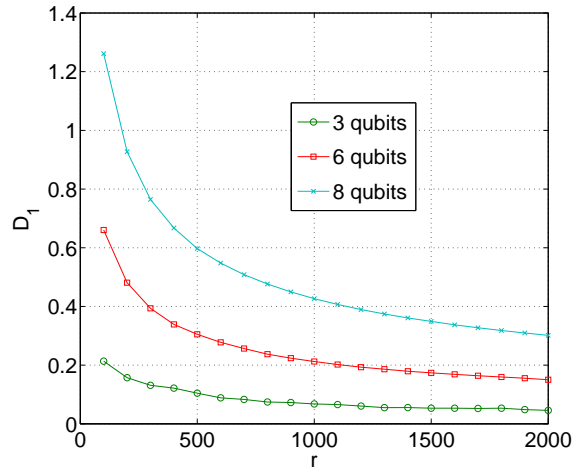} \caption{(Color online) Distance between empirical average and CUE exact
result ($\mathbb{I}/D^{N}$), as the number of states sampled is increased.
The size of the system is 3, 6 and 8 qubits, going from the bottom
to the top curve.}
\label{fig:D1-Micro}
\end{figure}

\begin{figure}[htp]
\centering{}\includegraphics[scale=0.4]{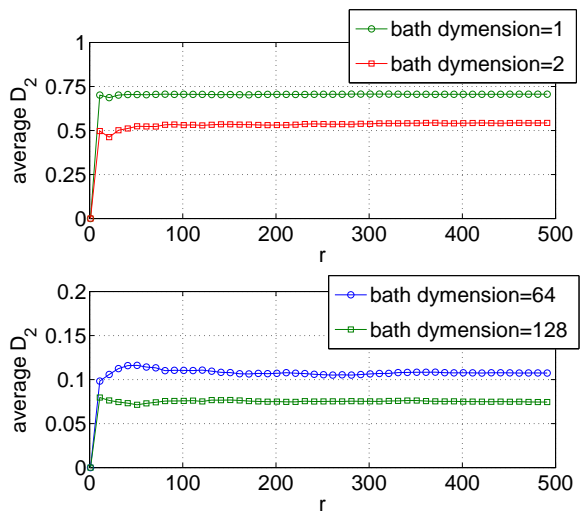} \caption{(Color online) 
Average Hilbert-Schmidt distance for a one-qubit
subsystem, $\overline{D_{2}}^{r}=\overline{||\rho_{s}-\overline{\rho}_{s}^{r}||}_{2}^{r}$,
as the number of states used is increased, and for different bath
sizes. }
\label{fig:Conv-D2}
\end{figure}

We first want to understand the effect of the number $r$ of sampled
states used in the averages.\textcolor{black}{{} }In Fig. \ref{fig:D1-Micro}
we evaluate the trace distance between the empirical average random
state and its exact value for the circular unitary ensemble (CUE)
which is the ensemble of unitary matrices corresponding to the Haar
measure, as a function of $r$:\textcolor{black}{{} $D_{1}\equiv\Vert\overline{\rho}^{r}-\mathbb{I}/D^{N}\Vert_{1}$.
The figure shows the results for different system sizes: 3, 6 and
8 qubits. As expected, it is observed that more states are needed
as the dimension of the Hilbert space is increased. At the level of
the subsystem, we also check how statistical fluctuations are suppresse}d
when increasing the number of sampled states. For this purpose we
consider the average Hilbert-Schmidt distance for a sub-system of
one qubit, $\overline{D_{2}}^{r}\equiv\overline{\Vert\rho_{s}-\overline{\rho_{s}}^{r}\Vert_{2}}^{r}$,
which is easier to evaluate numerically than the trace distance.\textcolor{black}{{}
}From Fig. \ref{fig:Conv-D2} we see that 500 states seem already
sufficient to suppress the statistical fluctuations in the subsystem.

\begin{figure}[htp]
\centering{}\includegraphics[scale=0.4]{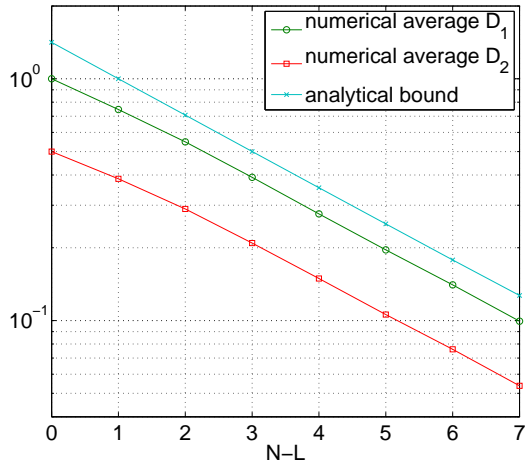} 
\caption{(Color online) Illustration of typicality for general states.
The figure shows the analytical value of the bound, $\sqrt{d_{S}/d_{B}}$
(with $d_{S}$ and $d_{B}$ the dimension of the
system and the bath), and the numerical average value of $\overline{D_{1}}^{r}$
and $\overline{D_{2}}^{r}$ for general random states distributed
with the Haar measure.}
\label{fig:4eps}
\end{figure}

Now, using 500 random pure states, we compare the bound given in \cite{PoShWi}
for $\overline{D_{1}}^{r}$ and its empirical value obtained from
the simulations. Fig. \ref{fig:4eps} shows how
the analytical value and the numerical results are pretty close and
scale in the same way with the bath size $n_{B}=N-L$. In the figure
we also plot the actual value of the average distance using the trace
norm and Hilbert-Schmidt norm.

\textcolor{black}{We now consider RMPS originated from random unitaries
appearing in the sequential generation scheme. The method used to
obtain random unitaries distributed according to the Haar measure
is documented in the literature \cite{Mez}. It makes use of the orthonormalization
of a random matrix where all the elements are i.i.d. Gaussian random
numbers of zero average and unit variance.}

\begin{figure}[htp]
\centering{}\includegraphics[scale=0.4]{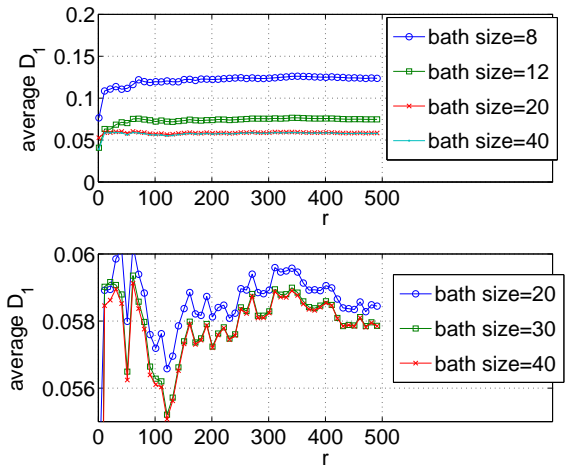} \caption{(Color online) Average trace distance from the average
state for a subsystem of one qubit, $\overline{D_{1}}^{r}=\overline{||\rho_{s}-\overline{\rho}_{s}^{r}||}_{1}^{r}$,
for homogeneous MPSs and OBC as the number of states used in the average
increases for $\chi=20$ and different bath sizes.}
\label{fig:Conv-MPS-OBC20}
\end{figure}

\begin{figure}[htp]
\centering{}\includegraphics[scale=0.4]{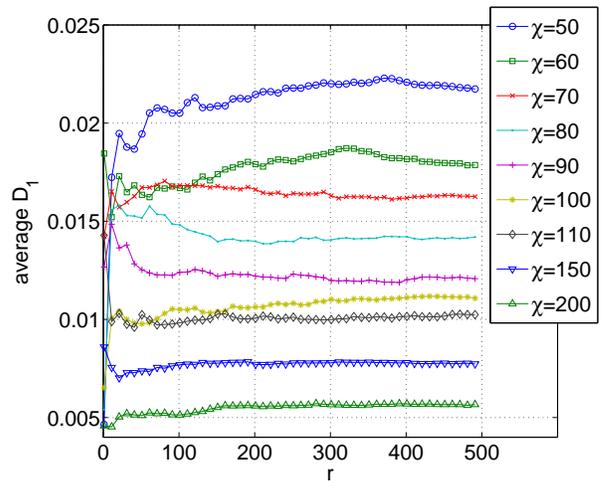} \caption{(Color online) Average trace distance from the average
state for a subsystem of one qubit, $\overline{D_{1}}^{r}=\overline{||\rho_{s}-\overline{\rho}_{s}^{r}||}_{1}^{r}$,
for homogeneous MPSs and OBC as the number of states used in the average
increases for $\chi=N-L=50,\,60,\,70,\,80,\,90,\,100,\,110,\,150\,\text{and}\,200$
(from top to bottom).}
\label{fig:Conv-MPS-OBC}
\end{figure}

Again we need to estimate the size $r$ of the sample, in order to
avoid statistical fluctuations that would be too strong. We look at
how the average trace distance from the average RMPS for the subsystem,
$\overline{D_{1}}^{r}\equiv\overline{||\rho_{s}-\overline{\rho_{s}}^{r}||_{1}}^{r}$,
converges when the sample size is increased. Fig. \ref{fig:Conv-MPS-OBC20}
shows this quantity for homogeneous MPSs with OBC, $\chi=20$ and
$n_{s}=1$. The different curves in the figure are for different bath
sizes: in the right we have the curves for baths of 20, 30 and 40
qubits (from top to bottom with 30 and 40 almost indistinguishable).
It may appear that for 500 states the average value has not totally
converged. However, if one compares more bath sizes, as done in the
upper part of Fig. \ref{fig:Conv-MPS-OBC20} with 8, 12, 20 and 40
qubits, it can be seen that the fluctuations are small in the scale
of interest. \textcolor{black}{This same qualitative behavior has
been observed for all the simulations we made, as illustrated in Fig.
\ref{fig:Conv-MPS-OBC}. Most of our computations shall therefore
consider a sampling from a set of 500 RMPS unless otherwise stated.}

\begin{figure}[htp]
\centering{}\includegraphics[scale=0.4]{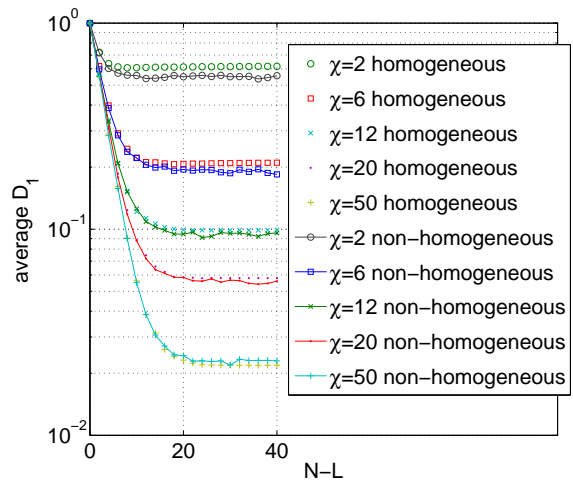} \caption{(Color online)
 $\overline{D_{1}}^{r}$ as a function of the
difference between the size of the system and the subsystem, for fixed
but different values of $\chi$ and for $L=1$. The symbols denote
homogeneous MPS, whereas the symbols joined by lines denotes non-homogeneous
MPS with the same value of $\chi$. Periodic boundary conditions are
imposed.}
\label{fig:LogD1-nb}
\end{figure}

\begin{figure}[htp]
\centering{}\includegraphics[scale=0.4]{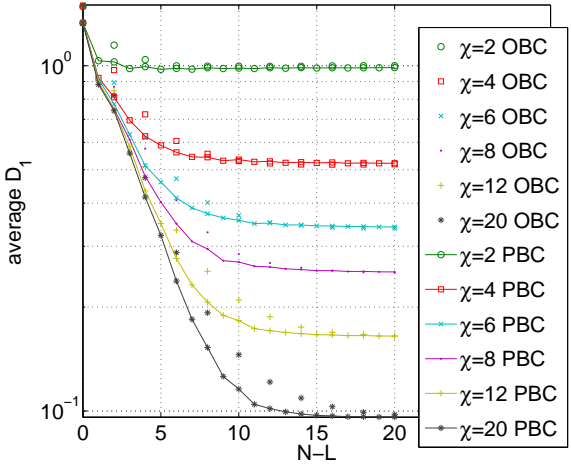} \caption{(Color online) $\overline{D_{1}}^{r}$ as a function of
the difference between the size of the system and the subsystem for
different values of $\chi$ and for $L=2$. The symbols
denote open boundary conditions, whereas the symbols joined by lines
denote periodic boundary conditions, both for homogeneous MPS.}
\label{fig:LogD1-nb-PBC}
\end{figure}

\textcolor{black}{In Fig. \ref{fig:LogD1-nb} we investigate the behavior
of $\overline{D_{1}}^{r}$ for non-homogeneous MPSs and homogeneous
MPSs with OBC. When we fix the value of $\chi$ and increase the number
of qubits in the bath, }$\overline{D_{1}}^{r}$\textcolor{black}{{}
starts to decrease, but soon reaches a constant value. This value
depend on $\chi$ and decreases as we increase $\chi$. Such behavior
is consistent with the previously mentioned analytic result \cite{GadeOZa}.
The same behavior is observed for larger subsystems (up to 5 qubits),
but with higher saturation values. For PBC the simulations are slower,
since we have a scaling with $\chi^{5}$, but we checked that until
values around $\chi=20$ the behavior is similar. This is shown in
Fig. \ref{fig:LogD1-nb-PBC}.}

\begin{figure}[htp]
\centering{}\includegraphics[scale=0.4]{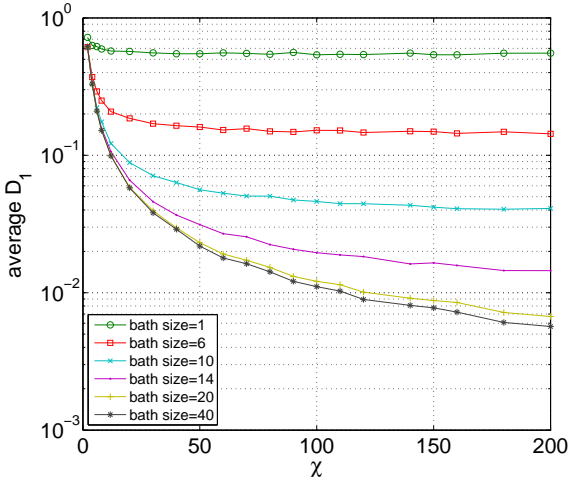} \caption{(Color online) $\overline{D_{1}}^{r}$ as a function of $\chi$
for different values of the bath's size $N-L$. We consider homogeneous
MPSs with OBC.}
\label{fig:LogD1-D}
\end{figure}

We now consider the dependence of $\overline{D_{1}}^{r}$ in $\chi$,
for fixed system and bath sizes. In Fig. \ref{fig:LogD1-D} it can
be seen that $\overline{D_{1}}^{r}$ also decreases with $\chi$,
until a value that depends on the bath size. For small system (up
to 8 qubits) we checked that this limiting value is the same as the
one for the CUE. Since $\chi$ can be viewed as a sort of correlation
length, this behavior is easily explained considering that when the
correlation length is of the order of the size of chain, $\overline{D_{1}}^{r}$
cannot decrease anymore.

\begin{figure}[htp]
\centering{}\includegraphics[scale=0.8]{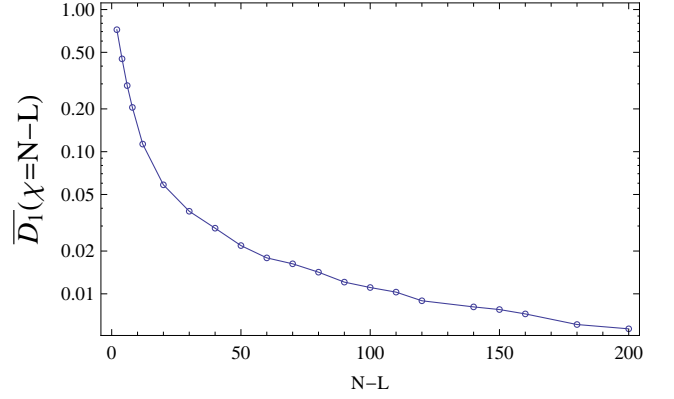} \caption{(Color online) $\overline{D_{1}}^{r}$ as a function of the
difference between the size of the system and the subsystem for $\chi=N-L$
. We consider homogeneous MPSs with OBC.}
\label{fig:LogD1-nb-D}
\end{figure}

\textcolor{black}{We now allow $\chi$ to scale linearly with the
size of the bath: $\chi=N-L$, see Fig. \ref{fig:LogD1-nb-D}. In
this case we observe that until $\chi=200$ the variance is decreasing
monotonically, which indicates that typicality can emerge already
for a linear scaling of $\chi$ with the number of particles.}

\subsection{Analytical Results for the average MPS}

An important aspect in the study of the concentration-of-measure
for RMPS is the characterization of the average state. Below we provide
an analytical expression for the averaged MPS state and numerical
verifications of the results obtained.

We use bold faced letters to denote vectors: ${\bf \mathbf{i}}\equiv(i_{1},\dots,i_{N})$.
Let \begin{equation}
|\psi\rangle=\sum_{\{\mathbf{i}\}}{\rm Tr}\left(A^{i_{1}}[1]A^{i_{2}}[2]\cdots A^{i_{N}}[N]\right)|i_{1}i_{2}\cdots i_{N}\rangle\end{equation}
 be an instance of RMPS, and the $A$-matrices are defined as $A^{i_{k}}\left[k\right]\equiv{\rm Tr}_{\mathcal{F}}\left[\left(|i_{k}\rangle\langle0|_{\mathcal{F}}\otimes\mathbb{I}_{\mathcal{A}}\right)\cdot U\left[k\right]\right],$
where $\mathcal{A}$ is the ancillary Hilbert space and $\mathcal{F}$
is the physical Hilbert space. For brevity of notation we will not
always specify the site index of the $A$-matrices and $U$-matrices.
The density matrix $\psi\equiv|\psi\rangle\langle\psi|$ is given
by \[
\psi(U)=\sum_{\{\mathbf{i},\mathbf{j}\}}{\rm Tr}\left(A^{i_{1}}\cdots A^{i_{n}}\right){\rm Tr}\left(A^{j_{1}}\cdots A^{j_{N}}\right)^{*}\]
\begin{equation}
\times|i_{1}\cdots i_{n}\rangle\langle j_{1}\cdots j_{n}|,\label{eq:aveDM}\end{equation}
\textcolor{black}{where $*$ denotes complex conjugation. The average
density matrix is given by }\begin{equation}
\overline{\psi}=\int_{{\rm Haar}}\psi(U)\,\, dU.\end{equation}
The factor in the coefficient of the density matrix can also be written
as \[
{\rm Tr}_{\mathcal{A}}\left(A^{i_{1}}\cdots A^{i_{N}}\right)=\]
 \[
{\rm Tr}_{\mathcal{A}^{\otimes N}}\left[S_{\mathcal{A}^{\otimes N}}\cdot(A^{i_{1}}\otimes\cdots\otimes A^{i_{N}})\right]=\]
 \begin{equation}
{\rm Tr}_{\mathcal{A}^{\otimes N}\otimes\mathcal{F}^{\otimes N}}\left[(|{\bf \mathbf{i}}\rangle\langle\mathbf{0}|_{\mathcal{F}^{\otimes N}}\otimes S_{\mathcal{A}^{\otimes N}})\cdot\bigotimes_{k=i_{1}}^{i_{N}}U_{k}\right],\end{equation}
where $S$ is the operator which cyclically permutes the states in
$\mathcal{A^{\otimes\mathrm{N}}}$ \begin{equation}
S|\alpha\rangle_{1}\dots|\alpha\rangle_{N}=|\alpha\rangle_{N}|\alpha\rangle_{1}\dots|\alpha\rangle_{N-1}.\end{equation}
 The coefficient of the density matrix (\ref{eq:aveDM}) can then
be rewritten as 
\begin{eqnarray*}
\lefteqn{ {\rm Tr}_{\mathcal{A}^{\otimes2N}\otimes\mathcal{F}^{\otimes2N}}
\left[
\left(
|\mathbf{i}\rangle\langle\mathbf{0}|\otimes S\otimes|\mathbf{j}\rangle\langle\mathbf{0}|
\otimes S
\right)
\times \right. }\\
&&\left(
\bigotimes_{k=1}^{N}U[k]
\right)
\otimes
\left.
\left(
\bigotimes_{k=1}^{N}U[k]^{*}
\right)
\right].
\end{eqnarray*}
 In a more compact form we can define a new averaged density matrix
in $\mathcal{F}^{\otimes2N}$ \begin{equation}
\overline{\Psi}\equiv Tr_{\mathcal{A}^{\otimes2N}}[(\mathbb{I}\otimes S\otimes\mathbb{I}\otimes S)\cdot\overline{\left(\bigotimes_{k=1}^{N}U[k]\right)\otimes\left(\bigotimes_{k=1}^{N}U[k]^{*}\right)}],\label{eq:average_state}\end{equation}
which is related to the previous $\psi$ in the following way \begin{equation}
\overline{\psi}_{\mathbf{i},\mathbf{j}}=\langle\mathbf{i},\mathbf{j}|\overline{\Psi}|{\bf \mathbf{0}},\mathbf{0}\rangle.\end{equation}
 In the previous expression the quantity \begin{equation}
\overline{\left(\bigotimes_{k=1}^{N}U[k]\right)\otimes\left(\bigotimes_{k=1}^{N}U[k]^{*}\right)}\end{equation}
carries information about the average over the Haar measure. Since,
for non-homogeneous RMPS, permutations of the factors inside the integral
are allowed, one sees that \begin{equation}
\overline{\left(\bigotimes_{k=1}^{N}U[k]\right)\otimes\left(\bigotimes_{k=1}^{N}U[k]^{*}\right)}=\otimes_{k=1}^{N}\overline{U[k]\otimes U[k]^{*}}=\Pi_{\chi D}^{\otimes N},\end{equation}
where $\Pi_{\chi D}$ is the density matrix of the maximally entangled
state, (see the appendix in \cite{ToGa}) \begin{equation}
\Pi_{\chi D}\equiv\frac{1}{\chi D}\sum_{l=1}^{\chi D}\sum_{l'=1}^{\chi D}|l,l\rangle\langle l',l'|.\end{equation}
 Now let us go back to the expression for $\overline{\Psi}$. Since
the operator $S$ that cyclically permutes the ancilla spaces acts
on $N$-tensor copies of the same state, it does not change the state
\[
Tr_{\mathcal{A}^{\otimes2N}}\left[(\mathbb{I}\otimes S\otimes\mathbb{I}\otimes S)\cdot\overline{\left(\bigotimes_{k=1}^{N}U[k]\right)\otimes\left(\bigotimes_{k=1}^{N}U[k]^{*}\right)}\right]\]
\begin{equation}
=Tr_{\mathcal{A}^{\otimes2N}}\left[\Pi_{\chi D}^{\otimes N}\right],\end{equation}
 and the trace will just restrict the projector to the maximally entangled
state over the physical subspace \begin{equation}
Tr_{\mathcal{A}^{\otimes2N}}\left[\Pi_{\chi D}^{\otimes N}\right]=\Pi_{D}^{\otimes N}.\end{equation}
 Now it is easy to see that \[
\overline{\psi}_{\mathbf{i},\mathbf{j}}=\langle\mathbf{i},\mathbf{j}|\Pi_{D}^{\otimes N}|{\bf \mathbf{0}},\mathbf{0}\rangle\]
 \begin{equation}
=\frac{1}{D^{N}}\sum_{{\bf \mathbf{l}}={\bf \mathbf{1}}}^{\mathbf{D}}\sum_{{\bf \mathbf{l^{\prime}}}=\mathbf{1}}^{\mathbf{D}}\langle\mathbf{i},\mathbf{j}|\mathbf{l},\mathbf{l}\rangle\langle\mathbf{l}^{\prime},\mathbf{l}^{\prime}|\mathbf{0},\mathbf{0}\rangle=\frac{1}{D^{N}}\delta_{{\bf \mathbf{i}},\mathbf{j}}.\label{eq:ave_mps}\end{equation}
 This proves that the average non-homogeneous RMPS is the completely
mixed state for any value of $\chi$ \begin{equation}
\overline{\psi}^{{\rm RMPS}}=\overline{\psi}^{{\rm CUE}}.\label{eq:result}\end{equation}
Since the average state is the same for the two ensembles, all the
functional depending only on the first moment of the distribution
will be identical in the non-homogeneous  RMPS ensemble and CUE ensemble.
This is the case for the expectation value of observables \begin{equation}
\overline{Tr\left(O\rho\right)}^{RMPS}=\overline{Tr\left(O\rho\right)}^{CUE}.\end{equation}

The case of homogeneous RMPS is more complicated, although a close
expression can be obtained starting from Eq. (\ref{eq:average_state}).
The problem comes from the fact that now it is not anymore possible
to factorize the average in the tensor product of unitaries, but nevertheless
the integral can be expressed as \cite{ToGa} \begin{equation}
\int_{Haar}\left(U\otimes U^{*}\right)^{\otimes N}dU=\sum_{\sigma}|\overrightarrow{P_{\sigma}}\rangle\langle\overrightarrow{P_{\sigma}}|,\end{equation}
where the vector $|\overrightarrow{P_{\sigma}}\rangle$ is obtained
superimposing the column of the matrix $P_{\sigma}$, an orthonormal
representation of the permutation $\sigma$ acting on $N$ elements.
In general it is hard to find such a representation for large $N$
and $D$.

\subsection{Numerical Results for the average MPS}

\begin{figure}[htp]
\centering{}\includegraphics[scale=0.45]{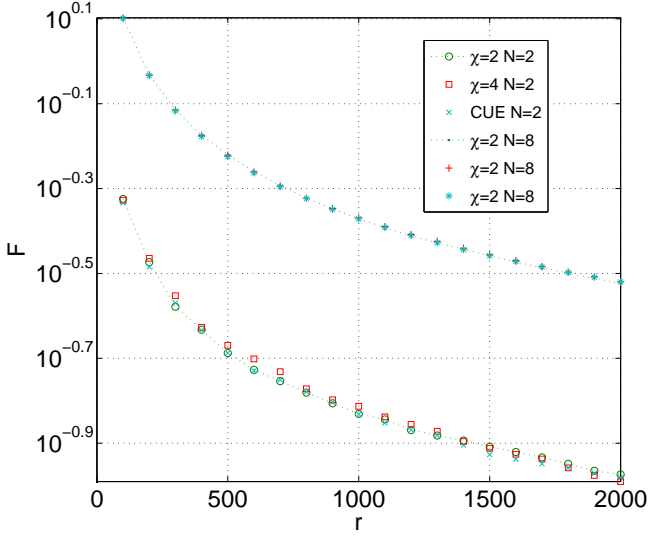} \caption{(Color online) The function $F\equiv\Vert\overline{\rho}^{r}-\mathbb{I}/D^{N}\Vert_{1}$
as the number of states used in the average is increased, for $N=2$
(bottom) and $N=8$ (top). For each system size we considered the
case of MPSs with $\chi=2$ and $\chi=4$ and general states (CUE),
however the different plots can barely be distinguished in the plot
scale. }
\label{fig:D1-Micro-MPS}
\end{figure}

\begin{figure}[htp]
\centering{}\includegraphics[scale=0.45]{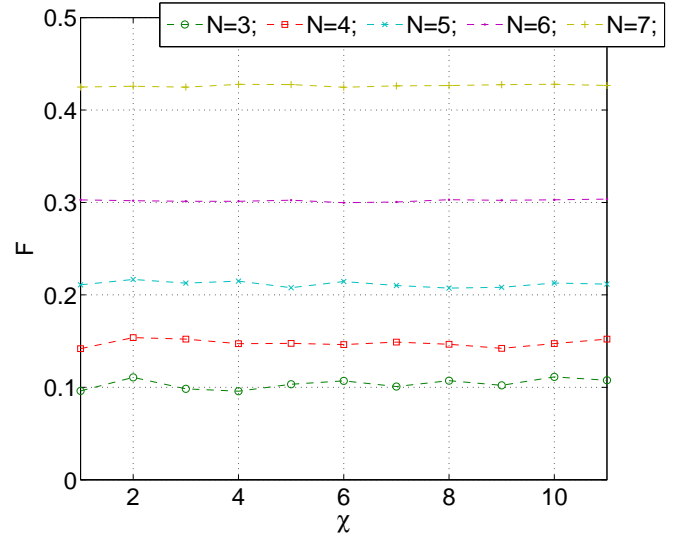} \caption{(Color online) The function $F\equiv\Vert\overline{\rho}^{r}-\mathbb{I}/D^{N}\Vert_{1}$
varying $\chi$ for different system sizes $N$. The numbers of states
used is $500$. It can be seen that there is no dependence on $\chi$
as expected from our analytical result in Eq. \ref{eq:result}. The
small fluctuations are due to the finite sampling.}
\label{fig:D1-Micro-Chi-MPS}
\end{figure}

In this section we numerically check the previous results.\textcolor{black}{{}
In Fig. \ref{fig:D1-Micro-MPS} we examine the distance between the
numerically averaged MPS state and its analytical value $\mathbb{I}/D^{N}$,
for systems of two and eight qubits, as the number $r$ of sampled
states increases. It can be seen that the average non-homogeneous
MPS converges to the mixed state as well as the average general state
(Fig. \ref{fig:D1-Micro}). The simulation shows that for these small
systems the average MPS state does not depend on $\chi$, as expected
from Eq. \ref{eq:result} (see also Fig. \ref{fig:D1-Micro-Chi-MPS}).
Note that this $\chi$-independence of the average MPS is a-priori
not at all obvious. }

Now we want to obtain more insight into the average MPS, studying
its behavior for larger system. We use the Hilbert-Schmidt norm between
the average MPS and the completely mixed state, since this can be
calculated efficiently, in contrast to the trace norm.

\begin{figure}
\centering{}\includegraphics[scale=0.4]{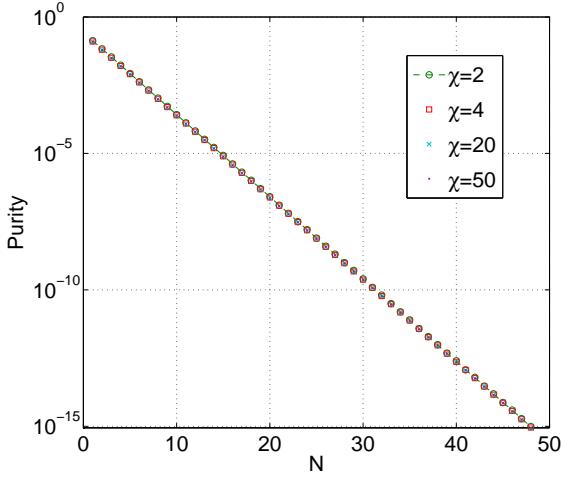}\caption{(Color online) Purity of the average MPS state, neglecting the
first term $1/r$ as a function of the number of qubits. The continous
line corresponds to $\frac{1}{2^{N}}$. The points are for values
of $\chi=2,\,4,\,20$ and $50$, but the difference can hardly be
seen.}
\label{fig:Purity}
\end{figure}

\begin{figure}
\centering{}\includegraphics[scale=0.45]{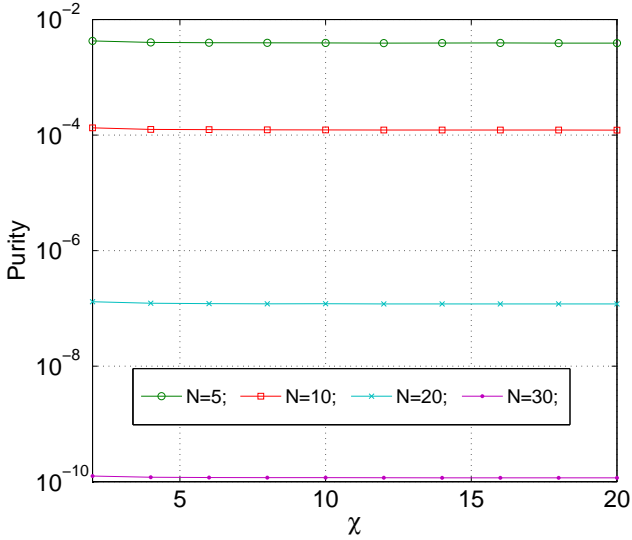}\caption{(Color online) Purity of the average MPS state $Tr\left((\overline{\rho}^{r})^{2}\right)$,
neglecting the first term $1/r$ as a function of $\chi$ for different
number of qubits $N$. There is no dependence on $\chi$ as expected.}
\label{fig:Purity-Chi}
\end{figure}

\begin{figure}
\centering{}\includegraphics[scale=0.45]{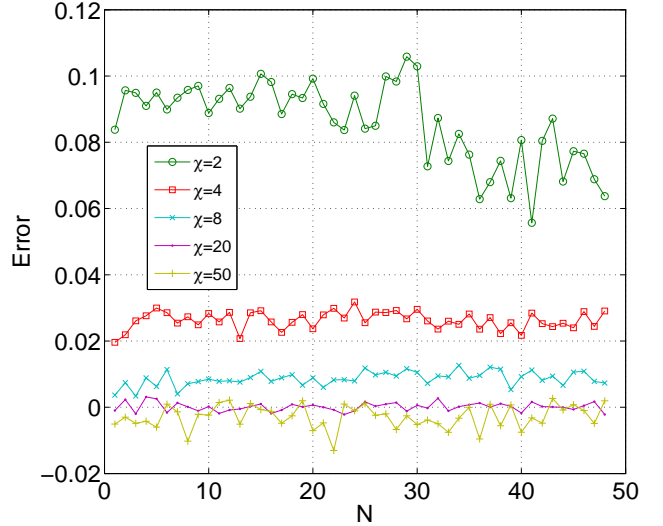}\caption{(Color online) 
Error in the purity of the average MPS state,
neglecting the first term $1/r$, as a function of the number of qubits
$N$ and for different values of $\chi$. Note that already for $\chi=2$
the error is around 10\% and does not increase with the system size.
The fluctuations are due to the finite size of the sample.}
\label{fig:ErrorPurity}
\end{figure}

Numerically, when we originate a number $r$ of RMPS $|\psi_{i}\rangle$
the ensemble average is given by \begin{equation}
\overline{\rho}^{r}=\frac{1}{r}\sum_{i=1}^{r}\frac{|\psi_{i}\rangle\langle\psi_{i}|}{|\langle\psi_{i}|\psi_{i}\rangle|}.\end{equation}
We then have \begin{equation}
||\overline{\rho}^{r}-\frac{\mathbb{I}}{D^{N}}||_{2}^{2}=\text{Tr}\left[\left(\overline{\rho}^{r}-\frac{\mathbb{I}}{D^{N}}\right)^{2}\right]=\text{Tr}\left[(\overline{\rho}^{r})^{2}\right]-\frac{1}{D^{N}}.\end{equation}
Is is clear from this expression that a necessary condition for a
small distance is a low purity of the average states. One may expect
that this limit is not reachable for mixed states constructed with
MPSs, since the entanglement of each element in the ensemble has a
bound depending on $\chi$. The purity of the average MPS is given
by \begin{equation}
\text{Tr}[(\overline{\rho}^{r})^{2}]=\frac{1}{r^{2}}\sum_{i,j=1}^{r}\frac{|\langle\psi_{i}|\psi_{j}\rangle|^{2}}{\langle\psi_{i}|\psi_{i}\rangle\langle\psi_{j}|\psi_{j}\rangle},\end{equation}
 and the overlap between different MPSs can be efficiently evaluated.
Before showing the results, let us rewrite the previous formula, remembering
that we want the purity to attain its minimum value $1/D^{N}$ \begin{equation}
\text{Tr}[(\overline{\rho}^{r})^{2}]=\frac{1}{r}+\frac{1}{r^{2}}\sum_{i\neq j}\frac{|\langle\psi_{i}|\psi_{j}\rangle|^{2}}{\langle\psi_{i}|\psi_{i}\rangle\langle\psi_{j}|\psi_{j}\rangle}.\end{equation}
 In the limit of a large sample, the first term vanishes, and the
second term should converge to $1/D^{N}$. It is clear that for large
systems, $D^{N}\gg r$, the principal limitation comes from the finite
size of the sample. Note that these expressions are also valid for
general states. Let us study how much the second term on the left
hand side of the previous equation differs from $1/D^{N}.$ This is
shown in Fig. \ref{fig:Purity} as a function of the system size,
and for different values of $\chi$. An exponential decrease of the
purity is observed for systems of up to 50 qubits. In the plot we
considered values of $\chi=2,\,4,\,20$ and $50$. However, they all
collapse to the same point. In fact, Fig. \ref{fig:Purity-Chi} shows
that there is no relevant dependence on $\chi$ with a very high accuracy.
We also study the relative error of the purity in relation to the
analytical value. This is an upper bound for the trace distance (again
the term $1/r$ is neglected):\begin{equation}
\frac{(\text{Tr}[(\overline{\rho}^{r})^{2}]-\frac{1}{r})-\frac{1}{D^{N}}}{\frac{1}{D^{N}}}=D^{N}||\overline{\rho}^{r}-\frac{\mathbb{I}}{D^{N}}||_{2}^{2}-\frac{D^{N}}{r}.\end{equation}
Fig. \ref{fig:ErrorPurity} shows this expression as a function of
the number of qubits and for different values of $\chi=2,4,8,20$
and $50$ (from top to bottom). Note that already for $\chi=2$ the
error is around 10\% and does not increase with the system size. Some
fluctuations are observed, and we believe they are due to the finite
size of the sample.

\subsection{Random MPS and uniformly distributed states}

The scheme for the sequential generation of RMPS can be viewed as
an algorithm for the efficient generation of random states. The computational
efficiency is given by the number of random unitary matrices needed,
which equals the size of the system, and by the size of these matrices,
which depends on $\chi.$ In this section we study how well the ensemble
of non-homogeneous  RMPS mimics a subset of statistical features associated
with the Haar measure. In particular we consider the average bipartite
entanglement, the minimum eigenvalue distribution and higher moments
of the reduced density matrix. 

\begin{figure}[htp]
\centering \includegraphics[scale=0.45]{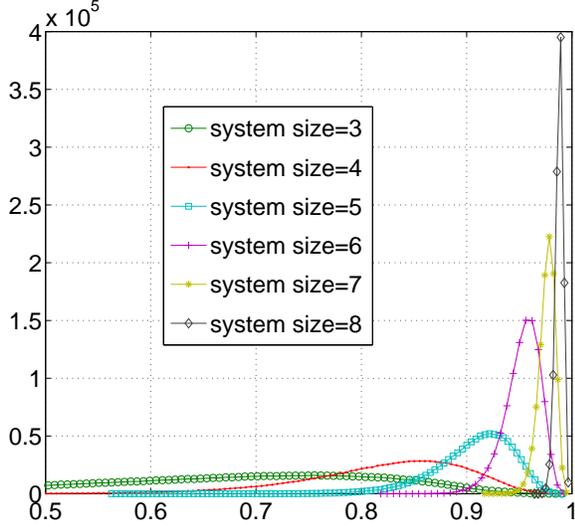} \caption{(Color online) Empirical probability distribution of the average bipartite entanglement
in RMPS.}

\label{fig:abe} 
\end{figure}

The distribution of entanglement produced by the present scheme for
the generation of random states can be a relevant quantity for quantum
information tasks. The average bipartite entanglement (ABE) or global
entanglement \cite{Brennen,MeWa}, here denoted with $Q,$ is a measure
of multipartite entanglement. It is defined as \begin{equation}
Q\equiv2-\frac{2}{N}\sum_{i=1}^{N}Tr[\rho_{i}^{2}],\end{equation}
 where $\rho_{i}$ is the reduced density matrix of the i-th qubit.
$Q$ can have values between 0 and 1, ranging from a product state
to a maximally entangled state. In Fig. \ref{fig:abe} we show histograms
of Q obtained over $10^{6}$ realizations of RMPS for different system
sizes. The distribution shows the same features as the ones obtained
with other random circuit schemes \cite{EmWeSa}. Increasing the dimension
of the Hilbert space the values of $Q$ concentrate around the average.
This is a manifestation of the concentration of measure phenomenon.

\begin{figure}[htp]
\centering \includegraphics[scale=0.4]{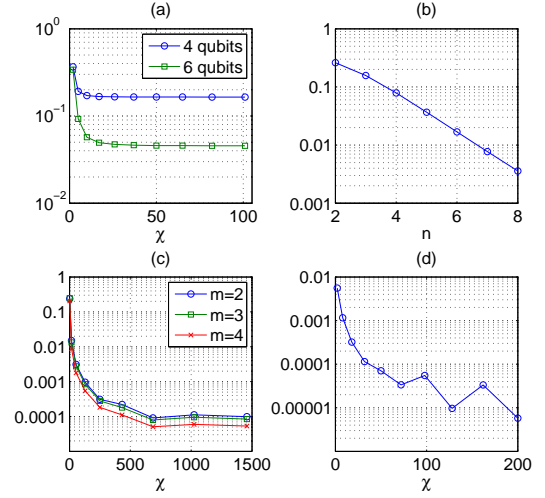} \caption{(Color online) (a) Plot of $|\overline{Q}^{RMPS}-\overline{Q}^{CUE}|$ as a function
of the RMPS rank $\chi.$ (b) Standard deviation of $Q$ for a system
of $n$ qubits. (c) $|\overline{Tr\rho^{m}}^{RMPS}-\overline{Tr\rho^{m}}^{CUE}|$
with $d_{A}=4$ and $d_{B}=16$. (d) $|\overline{\lambda_{min}}^{RMPS}-\overline{\lambda_{min}}^{CUE}|$
as functions of $\chi.$ }

\label{fig:avebe_4_6} 
\end{figure}

In Fig. \ref{fig:avebe_4_6}(a) we plot the difference between the
average value of $Q$ for RMPS and the exact value known for the CUE
ensemble: $|\overline{Q}^{RMPS}-\overline{Q}^{CUE}|$. The simulation
shows that the average $\overline{Q}^{RMPS}$, for $\chi$ sufficiently
large, depends only on the size of the system and the larger the system
the closer the average value is to the CUE exact result: $\frac{2^{n}-2}{2^{n}+1}$,
with $n$ the size of the system. Although from the numerical simulations
the difference between the two is always finite. This behavior is
similar to other random circuit constructions \cite{EmWeSa}, but
in our case the size of the system plays the role of the depth of
the random circuit and, for a fixed $\chi,$ determines the computational
cost of the simulation. This is also an indication of the fact that
the second moment of the RMPS $\overline{\rho\otimes\rho}^{RMPS}$
is different from $\overline{\rho\otimes\rho}^{CUE}$. This can be
seen from the fact that the average purity is a functional of two
copies of the random state: $\overline{Tr\left(\rho^{2}\right)}=Tr\left(S\cdot\overline{\rho\otimes\rho}\right)$,
where $S$ is the swap operator. The discrepancy between the RMPS
and CUE ensembles can then be detected from the second moment of the
distributions.

In Fig. \ref{fig:avebe_4_6}(b) we plot the decrease of the standard
deviation of $Q$ as a function of the system size $n$, with $\chi=64$.
The exponential decrease is again a signature of the concentration-of-measure
phenomenon for RMPS: for large system size almost all RMPS have $Q$
exponentially close to the average value $\overline{Q}^{RMPS}$.

The other quantities we want to compare with the ones obtained from
the CUE are the higher moments of the reduced density matrix evaluated
for a partition of the system in two parts, $A$ and $B$, of dimension
$d_{A}$ and $d_{B}$ respectively. Calculations in the CUE provide
the following exact results \cite{SoZy,Nec} \begin{equation}
\overline{Tr(\rho^{2})}^{CUE}=\frac{d_{A}+d_{B}}{d_{A}d_{B}+1},\end{equation}
 \begin{equation}
\overline{Tr(\rho^{3})}^{CUE}=\frac{d_{A}^{2}+3d_{A}d_{B}+d_{B}^{2}+1}{(d_{A}d_{B}+1)(d_{A}d_{B}+2)},\end{equation}
 \begin{equation}
\overline{Tr(\rho^{4})}^{CUE}=\frac{d_{A}^{3}+6d_{A}^{2}d_{B}+6d_{A}d_{B}^{2}+d_{B}^{3}+5d_{A}+5d_{B}}{(d_{A}d_{B}+1)(d_{A}d_{B}+2)(d_{A}d_{B}+3)}.\end{equation}
Fig. \ref{fig:avebe_4_6}(c) shows how the distance of the averaged
RMPS value from the CUE results depends on $\chi$. As can be seen
a finite and relatively small value of $\chi$ is sufficient to guarantee
a very good approximation of the CUE value. Again, after some point
there is no improvement in increasing the value of $\chi,$ as already
observed for the average bipartite entanglement in Fig. \ref{fig:avebe_4_6}(a).

We also consider the statistical properties of the minimum eigenvalue
of the reduced density matrix of a subsystem $A$ of dimension $d_{A}$.
This is a quantity related to the entanglement of the subsystem, and
its values can range from $0$, for product states, to $1/d_{A}$,
for maximally entangled states \cite{MaBoLa}. Fig. \ref{fig:avebe_4_6}(d)
shows the dependence on $\chi$ of $|\overline{\lambda_{min}}^{RMPS}-\overline{\lambda_{min}}^{CUE}|$,
where $\overline{\lambda_{min}}^{CUE}=1/d_{A}^{3}$ and $d_{A}=4$
(in a system of $6$ qubits). The figure indicates an exponential
convergence to the CUE value. The fluctuations seen in the plot are
due to the finite size of the sampling set and to the small value
of the quantity that we want to estimate (of the order of $10^{-5}$).
This result is again an indication of the good accuracy that can be
obtained in approximating some properties of the CUE with the RMPS
states.

\section{Conclusions}

In conclusion, in this work we have studied in detail a set of random
matrix product states (RMPS) introduced in Ref. \cite{GadeOZa}. As
already pointed out in that reference, RMPSs can be a useful tool
to address foundational problems of quantum statistical mechanics.
In particular here we have proved that the set of non-homogeneous
RMPS and the set of uniformly distributed general random states have
the same average state. This property, together with the validity
of the concentration-of-measure phenomenon, implies that any generalized
canonical state can be approximated by the reduced density matrix
of a random matrix product state, as long as the average random MPS
coincide with the associated averaged microcanonical ensemble. Let
us call $\Omega=\mathbb{I}_{R}/d_{R}$ the totally mixed state of
a global Hilbert space $\mathcal{H}_{R}$ satisfying some set of restrictions
denoted with $R$ (e.g. having a fixed energy), and whose MPS states
are denoted with $\psi_{R}\equiv|\psi_{R}\rangle\langle\psi_{R}|$.
Assuming that the average random MPS satisfies $\overline{\psi_{R}}^{MPS}=\Omega$
and substituting this in Eq.\ref{eq:com_state} it follows 

\begin{equation}
\textrm{Pr}\left[\Vert\rho_{s}-Tr_{Env}\Omega\Vert_{1}\geq4^{3L/2}\epsilon\right]\leq4^{L}c_{1}exp(-c_{2}\epsilon^{2}D\frac{\chi}{N^{2}}),\end{equation}
 with $\rho_{s}\equiv Tr_{Env}\psi_{R}.$ Since $Tr_{Env}\Omega$
is nothing but the generalized canonical state this prove our statement.
Notice that this is true only assuming measure concentration in the
restricted space and the equality $\overline{\psi_{R}}^{MPS}=\Omega$.
The present work focused on the case when the Hilbert space has no
restrictions. An interesting future direction of research would be
to check the identity between $\overline{\psi_{R}}^{MPS}$ and $\Omega$
when some kind of constraints are imposed. 

Another interesting application of RMPSs is in the field of pseudo-random
quantum circuits. We show that statistical properties of general quantum
random states, which are computational resources for some quantum
information tasks \cite{EmWeSa}, are very well approximated by RMPSs.
Since this states can be generated efficiently, as long as $\chi$
scales polynomially in the size of the system, they constitute an
efficient tool for the approximate simulation of random quantum states.

\smallskip{}

\begin{acknowledgments}
This work has been supported by NSF grants: PHY-803304,DMR-0804914.\end{acknowledgments}

\end{document}